# The influence of frequency and gravity on the orientation of active metallo-dielecric Janus particles translating under a uniform applied alternating-current electric field

Alicia Boymelgreen,*[a] Golak Kunti [b,c] Pablo Garcia Sanchez [d], Gilad Yossifon[e]

*Corresponding author. Email: aboymelg@fiu.edu*

a. *Department of Mechanical and Materials Engineering, Florida International University, Miami, Florida, USA, 33174.*

b. *Faculty of Mechanical Engineering, Indian Institute of Technology Delhi, New Delhi, India, 110016.*

c. *Faculty of Mechanical Engineering, Technion – Israel Institute of Technology, Technion City, Haifa, 3200003, Israel † Footnotes relating to the title and/or authors should appear here.*

d. *. Departamento de Electrónica y Electromagnetismo, Facultad de Física, Universidad de Sevilla, Avda. Reina Mercedes, Sevilla, Spain.*

e. *School of Mechanical Engineering, Tel Aviv University, Ramat Aviv 6997801, Israel*

## Abstract

Theoretical and numerical models of active Janus particles commonly assume that the metallo-dielectric interface is parallel to the driving applied electric field. However, our experimental observations indicate that the equilibrium angle of orientation of electrokinetically driven Janus particles varies as a function of the frequency and voltage of the applied electric field. Here, we quantify the variation of the orientation with respect to the electric field and demonstrate that the equilibrium position represents the interplay between gravitational, electrostatic and electrohydrodynamic torques. The latter two categories are functions of the applied field (frequency, voltage) as well as the height of the particle above the substrate. Maximum departure from the alignment with the electric field occurs at low frequencies characteristic of induced-charge electrophoresis and at low voltages where gravity dominates the electrostatic and electrohydrodynamic torques. The departure of the interface from alignment with the electric field is shown to decrease particle mobility through comparison of freely suspended Janus particles subject only to electrical forcing and magnetized Janus particles in which magnetic torque is used to align the interface with the electric field. Consideration of the role of gravitational torque and particle-wall interactions could account for some discrepancies between theory, numerics and experiment in active matter systems.

## Introduction

The field of active matter is an emerging area of interest with applications as broad as the fundamental understanding of non-equilibrium complex systems[1] to targeted drug delivery and diagnostics[2,3]. An active system is characterized by the ability of a single element to draw energy from its surroundings and convert it into mechanical motion. An isolated active object will translate independently along an individual path line, while the interaction of multiple elements will lead to collective motion such as swarming and flocking[4]. Such examples abound in nature and biology but more recently, interest has turned to synthetic active colloids whose mobility and interactions can be controlled through particle composition or geometry. In accordance with the scallop theorem[5], a key requirement for these particles to attain net motion is symmetry breaking at the particle level. One of the most commonly used symmetry-broken geometries is the Janus sphere – named after the Roman god of two faces due to their being comprised of two different chemical or material properties. For example, a Janus sphere where one hemisphere is coated in platinum while the other is an inert plastic will self-propel when placed in a solution containing hydrogen peroxide due to an asymmetric catalytic reaction at the platinum hemisphere[6]. External fields can also induce active motion, provided the propulsive mechanism is generated at the particle level[4].This motion is differentiated from 'phoresis' more commonly associated with external fields (e.g., dielectrophoresis, diffusiophoresis) wherein all elements are transported in a single direction under an externally applied gradient and the motion is not considered 'active'. Of specific relevance here is the motion of metallodielectric Janus spheres with one metallic and one dielectric hemisphere under uniform AC electric fields due to induced-charge electrophoresis (ICEP)[7] or self-dielectrophoresis (sDEP) depending on the frequency of the applied field[8].

The most common method of fabricating the Janus particles (JPs) to be used as synthetic active colloids is through deposition of a series of coatings over one hemisphere of commercially available, homogenous silica or polystyrene microparticles. The weight of this metal coating tends to cause the particles to rapidly sediment, indicating that the role of gravity is not in fact negligible even as it is commonly neglected in modelling due to the small size of the particles. Moreover, before swimming is initiated (either through the addition of the catalyst for self-propulsion or an applied field), the Janus sphere will orient with the metallic hemisphere facing downwards $(\alpha = 0)$ (Figure 1) signifying that the material nonuniformity of the Janus gives rise to a gravitational torque.

Upon activation, the particles tend to travel parallel to the wall in a quasi 2D system in which the interface between the hemispheres lies perpendicular to the substrate and the direction of motion. For electrokinetically driven colloids, this alignment, which acts against the gravitational torque arises from the electrostatic orientation of the induced dipole with the electric field as well as electrohydrodynamic torque generated from induced-charge electroosmotic (ICEO) flow around the metallic hemisphere[9–11] which propels the particle with the dielectric hemisphere forward. Both of these torques depend heavily on the frequency of the applied field. The proximity of the particle to wall, which breaks the symmetry in the vertical $x_2$ direction (Figure 1) will influence both the electrostatic and ICEO torques and could potentially introduce two additional contributions as well. Firstly, from a purely hydrodynamic perspective, a sphere translating at constant velocity near a wall will experience a torque due to the broken symmetry of the Stokes flow[12]. In the present case of electrokinetically driven Janus particles, since the direction in which the particle translates depends on the frequency of the electric field (dielectric hemisphere forward at low frequencies; metallic hemisphere forward at high frequencies)[8], this viscous torque can act either against (low frequencies) or in concert with (high frequencies) the gravitational torque. Additionally, at low frequencies, the particle-wall proximity also produces electrohydrodynamic (EHD) flow between the particle and the substrate[13] which could potentially result in a torque if the flow is asymmetric. However, since at the frequencies where this flow dominates (below the inverse of the charge relaxation of the powered electrodes) the JP's metallic hemisphere is fully shielded by the induced electric double layer at the particle surface, the wall induced EHD flow resembles that around a uniform dielectric sphere[13] and the torque contribution is small due to symmetry[8].

Previously, the roles of ICEO and dielectrophoretic (DEP) induced torques on the stable orientation angle have been analyzed numerically[10] and analytically [11] for the case of a Janus particle translating adjacent to an insulating wall. It was found that the Janus particle experiences a hydrodynamic attraction to the wall and is expected to tilt its interface at an angle $-45° < \alpha < 10°$ as it travels parallel to the wall in qualitative agreement with experimental observation[7]. In [14], it was demonstrated that a thick metallic coating could cause Janus particles to translate parallel – rather than perpendicular – to the electric field under induced-charge electrophoresis. In liquid crystals, it has been shown that the orientation of the metallodielectric Janus particles relative to the electric field is a function of frequency and voltage[15]. The relationship between equilibrium orientation and velocity was also studied analytically for a self-diffusiophoretic Janus sphere near a wall[16] in the absence of gravity, although numerical simulations indicate gravity will affect the equilibrium height above the wall and may induce transition between skimming and stationary states through the contribution to the stable angle of orientation[17,18]. A phase diagram comparing experimental and theoretical modelling of the sliding states is presented in [18]. A parametric study of the role of gravity on the equilibrium height for active Brownian particles with varying cap thicknesses is performed in [19], demonstrating that for realistic coatings $(>20nm)$ on microscale particles, gravity will dominate Brownian fluctuations and contribute to the orientation angle and equilibrium height.

In order to clearly demonstrate the role of the gravitational torque on the stable interfacial alignment of active colloids, we conduct a series of experiments in which the angle of orientation is measured for freely suspended Janus particles exposed to AC electric fields of varying frequencies. A second set of data in which a magnetic field is used to enforce alignment of the interface with the electric field is also obtained for comparison. These experiments are complemented by numerical simulations. It is readily apparent that although often neglected, gravity can play an important role in modulating the mobility of active colloids through the torque which makes a non-negligible contribution to the equilibrium orientation of the particle. Since the mobility of the particle is a function of the angle of orientation relative to the electric field[9], the gravitational torque also effectively modifies the particle velocity and can cause departure between experiments and theory - which generally considers the ideal case of the metallodielectric interface perfectly aligned with the electric field.

## Theoretical Modelling

To formulate the general problem, we begin by considering a freely suspended metallodielectric Janus sphere, of radius $a$ which is located at a distance $h$ (measured from the center of the particle) above a powered electrode (Figure 1). We define an inertial coordinate system $(x_1, x_2, x_3)$ aligned along and normal to the electrodes with origin at the center of the sphere. The electrode is subject to an oscillating ac electric field directed along the $x_2$ direction with an applied electric field $E(t) = \text{Re}\{E_0 e^{i\omega t}\}$, where Re $\{\ \}$ denotes the real part, $E_0$ is the magnitude of the applied field and $t$ is time. The spacing between the electrodes is given by a height, $H$.

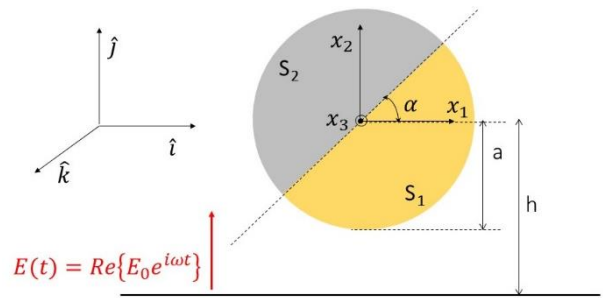

**Figure 1:** Schematic of Janus particle and coordinate system for theoretical model

Within this work, we focus on the steady state problem in which the particle is translating at a constant velocity, $U$ and the interface between the metallic (polarizable) and dielectric (insulating) hemispheres is aligned at an angle of $\alpha$ with respect to the AC electric field applied perpendicular to the channel walls, i.e., in the $x_2$ direction.

In the most general approach, where the particle is in close proximity to the wall, the equilibrium orientation of the particle is coupled to the velocity through the wall induced torque[10,11]. Following Goldman et al, [12], for a particle translating at a velocity $U$ in the $x_1$ direction, the rotational torque induced by wall proximity can be approximated as $\tau_s = 8\pi\eta a^2 U\left(-\frac{1}{10}\ln(h/a-1) - 0.1895\right)$ where $\eta$ is the dynamic viscosity of the liquid and for a particle translating in the direction of $x_1 > 0$, the torque (acting along the $x_3$ axis) acts in the clockwise direction.. For the maximum velocity in the current experiments, which is on the order of $16~\mu\text{m/s}$ and a hypothesized gap size of 150 nm corresponding to $h/a = 1.03$, subject to an applied voltage of magnitude $10~V_{pp}$, this corresponds to a non-dimensional torque $\tau_s/(\varepsilon E_0^2 a^3) \sim 0.01$ where $(\varepsilon = 80 \cdot 8.85 \cdot 10^{-12}~\text{F/m}, E_0 = 5/120 \times 10^{-6}~\text{V/m})$ which as will be shown in the ensuing section (Figure 3) is two orders of magnitude below the electrostatic and electrohydrodynamic torques. Accordingly, in the present analysis, the contribution of the wall-induced Stokes rotation is presumed negligible, thereby decoupling the equations of motion for the force and torque and significantly simplifying the problem. In this manner, the equilibrium orientation of the particle may be simply expressed as the zero sum of the electrostatic $\mathbf{\tau}_e^{(x_3)}$, electrohydrodynamic $\mathbf{\tau}_h^{(x_3)}$ and gravitational $\mathbf{\tau}_g^{(x_3)}$ torques (around the $x_3$ axis), i.e.,

$$\sum \mathbf{\tau}^{(x_3)} = \mathbf{\tau}_e^{(x_3)} + \mathbf{\tau}_h^{(x_3)} + \mathbf{\tau}_g^{(x_3)} = 0 \tag{1}$$

Focusing first on the electrostatic and electrohydrodynamic contributions, we follow our previous approach[8] and note that in the limit of "weak fields", where the electric potential $\phi$ is small with respect to the thermal scale, $\varphi_t = k_B T/ze$ (approx. 25 mV) (where $k_B$ is the Boltzmann constant, $T$ the absolute temperature, $z$ the valence and $e$ the elementary charge) the two governing equations may be linearized and decoupled [20]. The analysis is further simplified when the electric double layer (EDL) is much smaller than the particle radius such that the electrostatic potential within the electrolyte may be described by the Laplace equation

$$\nabla^2 \phi = 0. \tag{2}$$

The corresponding boundary conditions accounting for charging at the bottom powered electrode is given by

$$\left.\frac{\partial \phi}{\partial n}\right|_{x_2=-h} = i\Omega^*(\phi - V_0) \tag{3}$$

where $n$ is the normal, $\Omega^* = \omega a \lambda_0 / D$ in which $a$ is the particle radius, $\lambda_0$ is the length of the Debye layer and $D$ is the diffusivity of the electrolyte and $V_0 = E_0 H$ is the peak magnitude of the voltage difference ($H$ is the distance between the electrodes with the opposite electrode at $(x_2 = H - h)$ being grounded).

At the surface of the Janus sphere, a similar charging boundary condition is applied at the metallic hemisphere $(S_1)$ while the dielectric half $(S_2)$ is assumed to be perfectly insulating such that

$$\left.\frac{\partial \phi}{\partial n}\right|_{S_1} = i\Omega^*(\phi - \chi_0) \tag{4}$$

$$\left.\frac{\partial \phi}{\partial n}\right|_{S_2} = 0 \tag{5}$$

where $\chi_0$ represents the floating potential of the metallic coating determined by imposing zero total current on the particle surface, i.e., $\int \frac{\partial \phi}{\partial n} dS = 0$.

The time averaged electrical torque is obtained by integrating the Maxwell stress tensor $\mathbb{T}$ according to the following integration over any surface $S$ that encloses the JP

$$\mathbf{\tau}_e = \int_S \mathbf{r} \times (\mathbf{n} \cdot \mathbb{T}) dS = \hat{k} \int_S \left(x(\mathbf{n} \cdot \mathbb{T})_y - y(\mathbf{n} \cdot \mathbb{T})_x\right) dS \tag{6}$$

where $\mathbf{r}$ is the position vector of the points on the surface with respect to the center of the particle and $\mathbf{n}$ is a unit vector normal to it.

We note that in the limit that the particle is far from the wall, the time averaged electrical torque on the JP can be calculated from the induced dipole $\mathbf{p}$ as $\mathbf{\tau}_e = 1/2 \text{Re}\{\tilde{\mathbf{p}} \times \tilde{\mathbf{E}}^*\}$ where $\tilde{\mathbf{E}}^*$ is the complex conjugate of the electric field. The induced dipole phasor can be written as $\tilde{\mathbf{p}} = \alpha_\parallel \tilde{\mathbf{E}}_\parallel + \alpha_\perp \tilde{\mathbf{E}}_\perp$ where $\alpha_\parallel$ and $\alpha_\perp$ are the particle polarizabilities for an applied field parallel and perpendicular to the metallodielectric interface, respectively and may be calculated numerically from the solution to the electric potential for a particle subject to an applied field in that direction. $\tilde{\mathbf{E}}_\parallel$ and $\tilde{\mathbf{E}}_\perp$ are the parallel and perpendicular components of the electric field phasor, respectively. Accordingly, the time averaged electrical torque in the limiting case of the particle far from the wall is given by

$$\boldsymbol{\tau}_e = \frac{1}{2}\text{Re}\{\tilde{p}_\parallel \tilde{E}_\perp - \tilde{p}_\perp \tilde{E}_\parallel\}\hat{k} = \frac{1}{4}\text{Re}\{\alpha_\parallel - \alpha_\perp\}E_0^2 \sin(2\alpha)\hat{k} \qquad (7)$$

where $E_0$ is the magnitude of the applied field. It can readily be observed that for a particle in the bulk, the electrical torque is zero for $\alpha = \{-90°, 0, 90°, 180°\}$. Additionally, if $\text{Re}\{\alpha_\parallel - \alpha_\perp\} > 0$, the condition for stable equilibrium orientation is met for $\alpha = \{0, 180°\}$.

Once the solution to the electrostatic equation is found, the field induced fluid motion can be found from the solution to the unforced homogeneous dimensionless Stokes and the continuity equations,

$$\nabla P = \nabla^2 \mathbf{v}; \nabla \cdot \mathbf{v} = 0, \qquad (8)$$

(where $P$ denotes the hydrodynamic pressure, normalized with respect to $\varepsilon_0 V_0^2 / a^2$ where $\varepsilon_0$ is the permittivity of the solute and $V_0 = E_0 a$ while $\mathbf{v}$ is the solute velocity normalized by $\varepsilon_0 V_0^2 / \eta a$) through the application of the Helmholtz-Smoluchowski (H-S) slip velocity boundary conditions at the polarizable surface of the Janus sphere such that

$$\mathbf{v}|_{s_1} = -\frac{1}{4}\nabla_t |\phi - \chi_0|^2. \qquad (9)$$

where $\nabla_t$ corresponds to the gradient component tangential to the surface.

At the dielectric hemisphere, a no-slip condition is applied

$$\mathbf{v}|_{s_2} = 0. \qquad (10)$$

For simplicity, a no slip boundary condition is also applied at the upper $(x_2 = H - h)$ and lower $(x_2 = -h)$ electrodes. We note that previously it has been demonstrated that the proximity of the particle to the bottom electrode disturbs the uniformity of the local electric field, resulting in a tangential component that could drive electrohydrodynamic (EHD) flow along the substrate[13], the direction of which (injection or ejection from the "south pole" of the particle) depends on the particle polarizability. However, we have previously shown[8,21] that the low frequencies ($\Omega^* = \omega H^2/D \sim 1$) at which this EHD flow dominates also correspond with the regime in which the metallic surface of the Janus particle is fully screened by the induced EDL ($\Omega^* = \omega a \lambda_0/D \sim 1$) such that the electrostatic solution around the Janus particle represents that around a homogeneous dielectric sphere[9] and the EHD flow at the substrate corresponds to symmetric injection[13,21]. Accordingly, any potential contribution to the torque due to particle-induced electrohydrodynamic flow at the electrode will be zero due to symmetry[8].

The electrohydrodynamic torque is obtained by integrating the hydrodynamic stress tensor $\sigma_{ij}$ according to the following integration over the surface $S$ of the JP

$$\boldsymbol{\tau}_h = \int_S \mathbf{r} \times (\mathbf{n} \cdot \boldsymbol{\sigma}) dS = \hat{k}\int_S \left(x(\mathbf{n}\cdot\boldsymbol{\sigma})_y - y(\mathbf{n}\cdot\boldsymbol{\sigma})_x\right) dS. \qquad (11)$$

Finally, in order to consider the gravitational contribution to the stable orientation, we model the thin metallic coating which forms the polarizable hemisphere of the Janus particle as a surface mass density with the following value

$$\bar{\rho}_s = \frac{dM}{dS} = \sum_{i=1}^n (\rho_i - \rho_w) t_i, \qquad (12)$$

where $n$ is the number of coatings and $\rho_i$ is the volume density of the material and $t_i$ the corresponding coating thickness. The effective immersed mass density is given by $\bar{\rho}_i = \rho_i - \rho_w$ where $\rho_w$ is the density of the electrolyte. The gravitational torque with respect to the center of the sphere $(r = 0)$ is calculated as

$$\boldsymbol{\tau}_g = \int \mathbf{r} \times d\mathbf{F} = \int_S (\mathbf{r} \times \mathbf{g}) \bar{\rho}_s dS. \qquad (13)$$

where $\mathbf{g} = -g\hat{j}$ is the gravitational acceleration and $d\mathbf{F} = \bar{\rho}_s \mathbf{g} dS$. Using spherical coordinates yields the gravitational torque in terms of $\bar{\rho}_s$, $a$ and $\alpha$

$$\boldsymbol{\tau}_g = \bar{\rho}_s a^3 g \int_0^\pi \int_{\alpha-\pi}^\alpha [\cos\theta \sin\theta\, \hat{i} - \sin^2\theta \cos\varphi\, \hat{k}] d\varphi d\theta = -\pi\bar{\rho}_s a^3 g \sin\alpha\, \hat{k}. \qquad (14)$$

The torque contributions from Eqs.(6), (11) and (14) are substituted into Eq.(1) to estimate the equilibrium angle of orientation in the limit that the particle is translating sufficiently slow and far from the wall such that rotation induced by viscous drag near the wall may be neglected.

## Numerical Simulations

Numerical simulations are performed in 3D in COMSOL multiphysics v5.3 for the governing Laplace (2) and Stokes and continuity (8) equations. The domain is cylindrical with a radius of $50a$ and height $H = 24a$, where $a$ is the radius of the Janus particle (Figure 1). The particle is positioned at a distance $h$ (measured from the centre of the particle) above a powered electrode subject to an applied voltage $V(t) = \text{Re}\{V_0 e^{i\omega t}\}$; $V_0 = 1$ for ease of normalization. In accordance with the fitting demonstrated in [8] we have approximated the typical value of $h/a = 1.03$ corresponding to a gap width of $150\ \text{nm}$ which is also in accordance with the thin EDL approximation. The value of $\alpha$ is varied from $0$ to $90°$. The electrostatic boundary conditions at the bottom electrode is given by Eq. (3) and the upper electrode is grounded. A no slip boundary condition is applied at both the upper and lower electrodes. At the JP surface, the electrostatic boundary conditions are defined by Eqs.(4)-(5) and hydrodynamic by Eqs. (9)-(10). At all other walls, an insulating and no-slip boundary condition is applied to the electrostatic and hydrodynamic equations respectively.

## Experimental Methods

### JP and Chip fabrication

The Janus particles were manufactured following the methodology in Ref.[21]. Briefly, 10 µm polystyrene (Ps) particles (Fluro-max) were half coated with a thickness of 15 nm chrome for adhesion followed by 50 nm of ferromagnetic nickel and then 15 nm of gold. The Janus particles were magnetized by placing the substrates between two neodymium magnetic blocks (14 x 12 x 19 mm in size) with opposite poles facing each other. The JPs were sonicated in DI water for release. Prior to the experiments, the JPs were suspended in KCl electrolyte ($5\times10^{-5}$ M), a small quantity (0.05%) of nonionic surfactant (Tween 20 (Sigma-Aldrich)) was added to the electrolyte in order to minimize adhesion to the ITO substrate (measured conductivity 7 µS/cm).

The Janus particles were placed in a microfluidic chamber sandwiched between two parallel plate electrodes comprised of Indium tin oxide (ITO) coated glass slides (Delta Technologies)) (Figure 2a,b). The bottom electrode was coated with 50 nm of silicon dioxide ($SiO_2$) to avoid surface adsorption of JP. The chamber is formed by a silicon spacer of 120 µm thickness (Grace Bio). Upon placement in the experimental cell, the JPs sank to the bottom substrate. To avoid evaporation, two holes are drilled into the top ITO substrate and a reservoir pasted around the holes to store electrolyte. For the orientation of the JP at a specific angle a neodymium magnet block ($14 \times 12 \times 19$ mm in size) was placed at a distance of 7.5 cm. The large distance between the magnet and the chamber results in an approximately uniform magnetic field while avoiding magnetophoresis; verified by the lack of JP motion in the presence of the magnet. The electrodes were connected to a signal generator (Agilent 33250A) and the signal was monitored by an oscilloscope (Tektronix TPS-2024). Images were taken with an Andor Neo sCMOS camera attached to a Nikon TI inverted epi-fluorescent microscope with x20 objective lens.

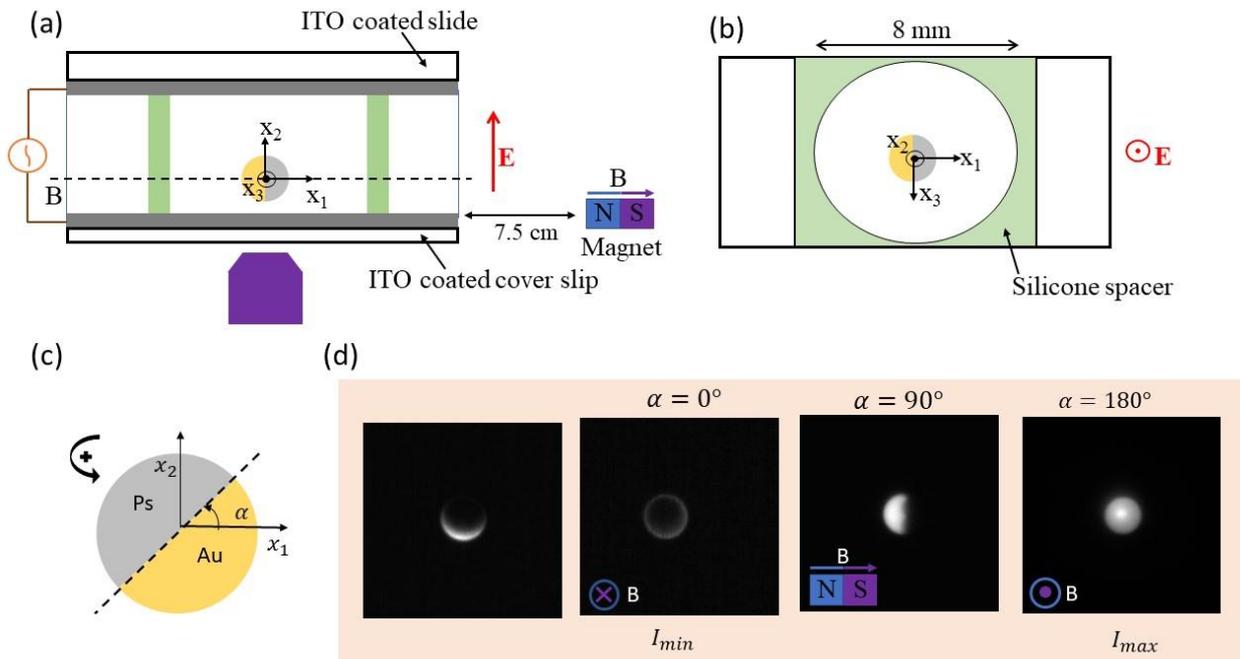

**Figure 2:** Experimental setup and analysis (a) and (b) Schematic illustration of experimental setup. (c) Schematic side view ($x_1$-$x_2$ plane) of the JP defining the orientation angle and direction of torque (d) Different orientations by applying magnetic field without electric field. The bright side corresponds to the fluorescent Ps while the dark hemisphere is Au coated

**Image analysis**

The velocity of the Janus particles was measured by automated tracking using the Speckle Tracker plugin[22] in ImageJ software. The orientation was obtained through measuring the fluorescence intensity of the images according to [23]

$$\alpha = 2\sin^{-1}\left(\sqrt{I^*}\right), \tag{15}$$

where $\alpha$ is the angle of orientation of the interface relative to electric field (Figure 2c). $I^*$ is the nondimensional areal integrated intensity of the Janus particle obtained by normalizing the measured intensity $(\bar{I})$ with the extreme cases of a particle facing with the metallic $(I_{min})$ and dielectric $(I_{max})$ hemispheres downwards (Figure 2d) such that

$$I^* = \frac{\bar{I} - I_{min}}{I_{max} - I_{min}}. \tag{16}$$

The reference measurements of $I_{min}$ and $I_{max}$ were taken with the magnet activated as illustrated in Figure 2d.

## Results and Discussion

The contributions of electrostatic and hydrodynamic torques as a function of frequency are obtained from the numerical simulations according to Eqs.(6) and (11) and compared with the gravitational torque calculated from Eq.(14) for a JP $(a=5\ \mu m)$ coated with a layer of chrome $(t_1=15\ nm, \rho_1=7.15\cdot 10^3\ kg/m^3)$, nickel $(t_2=50\ nm, \rho_2=8.9\cdot 10^3\ kg/m^3)$ and gold $(t_3=15\ nm, \rho_3=19.32\cdot 10^3\ kg/m^3)$ such that $\bar{\rho}_s = 0.762\cdot 10^{-3}\ kg/m^2$ in Figure 3. All torques are normalized by $\varepsilon a^3 E_0^2$. We note that since all torques are scaled by $a^3$ such that within the thin film approximation, the torque depends only on the thickness of the metallic coating and not the particle size. The grey shaded area indicates the region of unstable orientations while the intersection of the absolute value of the gravitational curve (green) with the total (purple) of the electrostatic (blue) and hydrodynamic (orange) torques represent the equilibrium orientation angle for a specific applied voltage.

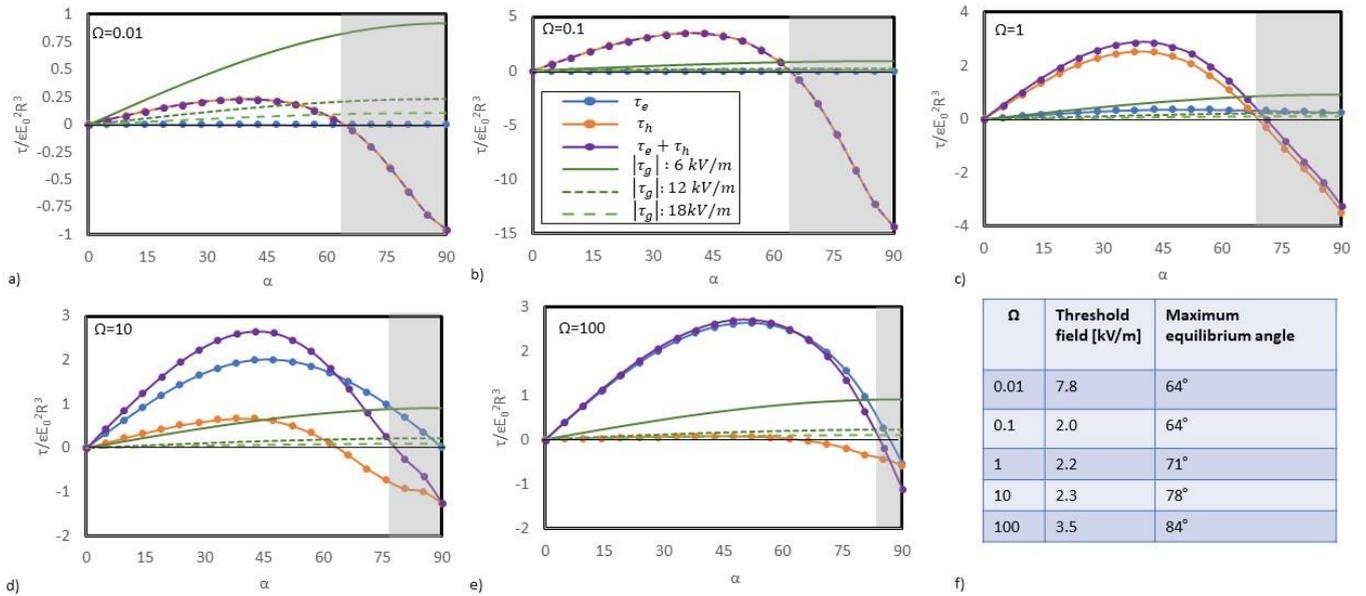

**Figure 3**:(a-e) Plot of electrostatic (blue), electrohydrodynamic (orange) and their superposition (purple) induced torques with the absolute value of the gravitational (green) torque for varying frequencies of the applied field: a) Ω=0.01, b) Ω=0.1, c) Ω=1, d) Ω=10 e) Ω=100. The grey shaded region indicates the region of unstable angles. The absolute value of the (negative, i.e. clockwise as depicted in Fig.2c) gravitational torque has been plotted to illustrate the intersection of gravitational curves (green) with the sum of the electrostatic and electrohydrodynamic torques (purple) representing stable orientation angle for a specific applied voltage. f) Summary of threshold field and stable equilibrium range for varying Ω.

We first note that the relative roles of the torques vary with the frequency of the applied field with the hydrodynamic torque - reflecting the induced-charge electroosmotic contribution - reducing with increasing frequency as expected when the inverse of the charge relaxation time of the particle is exceeded, i.e., $\Omega > 1$. The role of the magnitude of the applied voltage is made evident through the scaling of the gravitational torque such that its influence is reduced as the voltage is increased. Accordingly, at low frequencies, for small applied fields (e.g. $\Omega = 0.01, E_0 = 6$ kV/m in Fig.3a) – we observe that there is no equilibrium angle between 0 and 90 degrees and thus the particle will remain with its metallic hemisphere facing downwards ($\alpha = 0°$). Thus, by extension, there exists a threshold field – which is a function of frequency – below which the particle will not respond to the electric field and will remain in the unforced equilibrium position of $\alpha = 0°$ (Figure 3f).

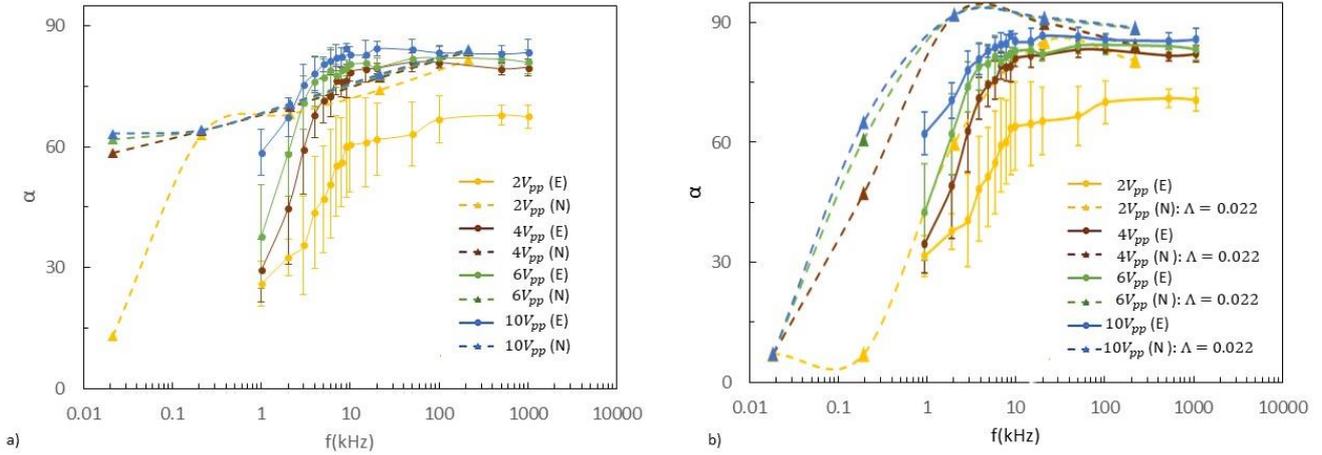

**Figure 4:** Measurement of stable orientation angle as a function of frequency for varying applied fields. a) experimental (solid line) and numerical (dashed line) values, b) experimental values with numerical model considering suppression of ICEO velocity by a factor of $\Lambda = 0.022$.

Additionally, we note that as the frequency and voltage of the applied field increase, the value of the equilibrium angle increases such that the particle interface becomes more closely aligned with the electric field.

In Figure 4a-b, we plot the experimentally measured value of $\alpha$ (Eq.(15)) as a function of frequency for varying voltages. We compare the experimental values with the numerical simulations where the orientation angle is extracted by finding the intersection of the absolute value of the gravitational torque with the sum of electrostatic and hydrodynamic torques for the prescribed voltage. The present model is able to qualitatively capture the key trends including the decrease of $\alpha$ with increasing voltage and frequency. Quantitatively, we note maximum discrepancy in the low frequency limit where ICEO is dominant. The experimental suppression of induced-charge electrokinetic flows, which would also reduce the associated torque has been widely observed experimentally and attributed to a number of sources including the capacitance of the Stern layer[24] and high voltage effects[25]. According to the former, the presence of the Stern layer is expected both suppress the experimental velocity and shift the frequency dispersion which are multiplied by a factor of $\Lambda = 1/(1+\delta)$ where $\delta = C_d/C_s$ is the ratio of the capacitance of the Debye $(C_d)$ and Stern $(C_s)$ layers respectively[24]. In order to illustrate the effect that this suppression could have on the equilibrium angle, we compare the experimentally maximum measured ICEP velocity at 10 $V_{pp}$ (dotted blue line, Figure 5b) at 1kHz with the theoretical velocity[8] to obtain a fitting factor of $\Lambda = 0.022$ which is applied to the magnitude of the ICEO torque $(\tau_h)$ such that $\tau_h|_{\Lambda=0.022} = \Lambda \tau_h$. The adjusted equilibrium values of $\alpha$ are plotted in Figure 4b. It is evident that in accordance with experimental observation, the suppression of ICEO would increase the value of α at low frequencies as well as the rate of change for moderate frequencies in qualitative agreement with experimental observation. We note that although the Stern layer capacitance will also cause a shift in the frequency distribution[24], which is also evident in the current experiments, we have not included that effect in the fitting for Figure 4b since there exists a nonlinear coupling between the capacitance of the EDL at the metallic hemisphere of the JP (which controls the ICEO flow) and the capacitance of the $SiO_2$ coating and EDL at the ITO surface which cannot be accounted for with this simplified model[8]. This coupling would also contribute to the suppression of the electrostatic torque $(\tau_e)$ which correlates with the overprediction of $\alpha$ at high frequencies. Full quantitative agreement, accounting for the coupled ICEO suppression and frequency shift of electrostatic and electrohydrodynamic torques due to the presence of the $SiO_2$ coating as well as the particle height which can affect both magnitude and frequency dispersion [8] is left for future work.

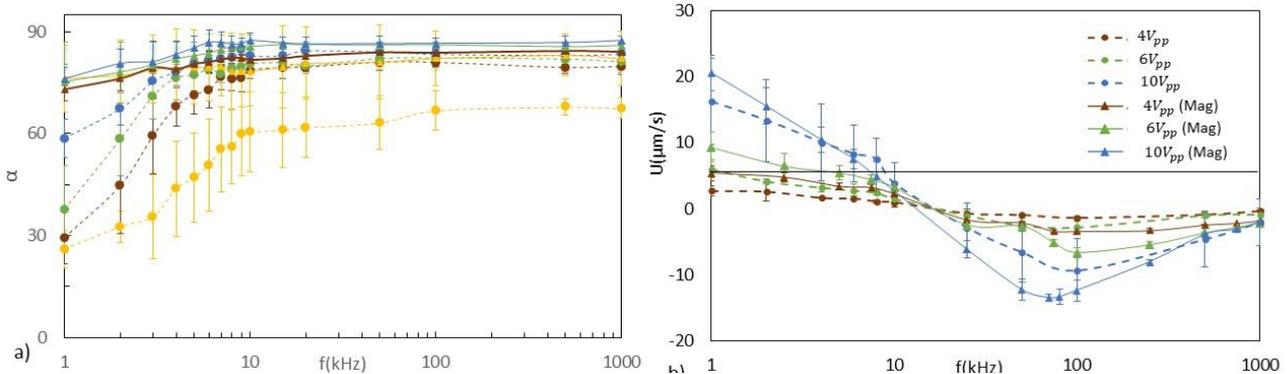

**Figure 5:** Comparison of equilibrium angle (a) and velocities (b) with (solid line) and without (dotted line) an applied magnetic field as a function of frequency for varying voltages.

Practically, one important outcome of the gravitational torque - and the resultant increased equilibrium value of $\alpha$ - is the corresponding reduction in velocity as the angle departs from the optimum value of 90° commonly assumed in theoretical and numerical simulations[9]. To illustrate this phenomenon, we compare the velocity and angle of orientation of a freely suspended JP with one on which a magnetic field is applied to align the metallodielectric interface with the electric field (in the $x_2$ direction) and perpendicular to the substrate and direction of motion. As expected, in the presence of the magnetic field, the measured angle of orientation is closer to 90° at all frequencies and the velocity is increased as compared to the freely suspended sphere (Figure 5). We note that the onset of the saturation of the value of $\alpha$ to a constant value (Figure 5a) at ~10 kHz, corresponds with the regime where the particle switches direction to travel under sDEP (Figure 5b). This is in accordance with the fact that beyond the charge relaxation time of the particle, the EDL cannot form to screen the metallic hemisphere and thus both ICEP decays to zero and the torque balance is dominated by the interplay between gravitational and electrostatic contributions. Interestingly at low frequencies, where the JP velocity is at a maximum, the stable orientation under the applied magnetic field has a maximum value of $\alpha \sim 75°$. This departure from alignment with the electric field could reflect the large negative ICEO torques at small frequencies and values of $\alpha$ (Figure 3) which co-act with the gravitational torque against the magnetic forcing.

## Conclusions

The present work illustrates that although often assumed negligible in microfluidic and soft matter systems, gravity can play an important role in modulating the stability and mobility of active colloids. Specifically, it is shown that the asymmetric weight distribution of a metallodielectric sphere generates a gravitational torque that can oppose the torques induced by the driving mechanisms of active colloids which tend to align the interface perpendicular to the direction of motion. For the specific case of electrokinetically driven colloids translating near a wall, we use a combination of experiment and numerics to show that the gravitational torque is counteracted by an electrostatic and electrohydrodynamic torque. Since the magnitude of these field induced torques are a function of frequency and voltage, we observe that the equilibrium angle of orientation likewise varies with these parameters, with the maximum angle between the interface and electric field obtained at low voltages where gravity dominates and low frequencies characteristic of induced charge electroosmosis.

A direct outcome of the departure from the standard theoretical model of a particle with an interface parallel to the electric field is a reduction in particle mobility which is demonstrated here through comparison of the velocity of freely suspended spheres and those whose interface are maintained in alignment with the electric field using magnetic forcing. Thus it is apparent that the effect of gravitational torque on angle of orientation could partially explain discrepancies between experimental results and theoretical/numerical models which tend to assume the particle translates with its interface aligned with the electric field. We note that the present model assumes a uniformly thin metallic coating; future work examining the effect of coating thickness on the equilibrium angle, accounting for variations in both weight distribution[18,26] and polarizability[14] could provide further insight on the interplay between gravitational and electrostatic torques.

Recognizing the role of gravity in soft matter systems is not only important in terms of bridging discrepancies between experiment and theory but is critical to the development of 3D active systems. For example, one emerging approach towards breaking out of the quasi 2D systems in which particles are confined to the plane immediately above the substrate, is the intelligent design of active colloids which are capable of translating in helical pathways either due to geometric features [27,28], patterning of the active patch (most commonly metal)[29,30] or a combination of these. Consideration of the gravitational torque arising from a nonuniform mass distribution could be critical to the practical realization of translation in three dimensions. Moreover, beyond single particle mobility, the angle of orientation will also affect particle-particle interactions in concentrated active systems which will in turn modulate the patterns of collective motion.

## Conflicts of interest

There are no conflicts to declare.

## Acknowledgements

AB acknowledges the support of NSF-CASIS #2126479. GY acknowledges the support of ISF 1934/20. PGS acknowledges financial support by ERDF and Spanish Research Agency MCI under contract PID2022-138890NB-I00.